# Railroad Bailouts in the Great Depression

2 authors:

Lyndon Moore
Monash University (Australia)
22 PUBLICATIONS   167 CITATIONS

SEE PROFILE

Gertjan Verdickt
KU Leuven
18 PUBLICATIONS   56 CITATIONS

SEE PROFILE

Some of the authors of this publication are also working on these related projects:

Historical Life Insurance  View project

The Great Depression  View project

# Railroad Bailouts in the Great Depression*

LYNDON MOORE  
*Monash University*

GERTJAN VERDICKT  
*KU Leuven*

**Abstract**

The Reconstruction Finance Corporation and Public Works Administration loaned 50 U.S. railroads over $1.1 billion between 1932 and 1939. The government's goal was to decrease the likelihood of bond defaults and increase employment. Bailouts had little effect on employment, instead they increased the average wage of their employees. Bailouts reduced leverage, but did not significantly impact bond default. Overall, bailing out railroads had little effect on their stock prices, but resulted in an increase in their bond prices and reduced the likelihood of ratings downgrades. We find some evidence that manufacturing firms located close to railroads benefited from bailout spillovers.

* We thank Sriya Anbil, Nabil Bouamara, Fabio Braggion, James Brugler, Viet Cao, Julio Crego, Abe de Jong, Toby Daglish, Hans Degryse, Geert Dhaene, John Friedman, Rik Frehen, Carola Frydman, Neal Galpin, Will Goetzmann, Florian Hoffmann, Peter Koudijs, Joseph Mason, Chris Meissner, Stefan Nagel, Sebastian Pfeil, Magdalena Rola-Janicka, Claus Schmitt, Yulong Sun, Stijn van Nieuwerburgh, Marno Verbeek, Angela Vossmeyer, and Barry Williams for helpful comments. We are grateful to participants at the Australian National University, CEPR International Macro History Online Seminar, Financial History Workshop (Antwerp), Irish Academy of Finance, KU Leuven, Latrobe University, LSE-UC Davis Economic History Seminar, Massey University, Monash University, Oslo Metropolitan University, Paris School of Economics, Southern Economic Association (New Orleans), Tilburg University, Queen's University Belfast, University of Antwerp, University College Dublin, University of Melbourne, and World Cliometric Workshop (Dublin) for helpful suggestions. Hannah Merki, Karen Needles, and Enzo Peeters provided excellent research assistance. We thank Gustavo S. Cortes for sharing his data.

Lyndon Moore (corresponding author), lyndon.moore@monash.edu  
Gertjan Verdickt, gertjan.verdickt@kuleuven.be

**1. Introduction**

The Reconstruction Finance Corporation (RFC) was created by President Hoover in early 1932 during the depths of the Great Depression. The objective of the RFC was to "make temporary advances upon proper securities to established industries, railways, and financial institutions which cannot otherwise secure credit, and where such advances will protect the credit structure and stimulate employment" (hereafter referred to as "the twin objectives").[1] The Corporation approved $3.9 billion in loans from 1932 until 1939.[2] We call such loans 'bailouts' because they were provided at below-market interest rates to companies that struggled to access credit from commercial sources.[3] Although most of these loans went to the financial sector (e.g., Mason (2001), Calomiris, Mason, Weidenmier, and Bobroff (2013), and Butkiewicz (1995, 1997)), railroads were, by far, the largest non-financial sector recipient (i.e., $1.17 billion, including rollovers) during this eight-year period. In this paper, we explore how the RFC's bailouts for railroads, along with limited assistance from the Public Works Administration (PWA), impacted the economy.

First, we study the impact of the bailouts on the assisted railroads themselves, given the RFC's twin objectives. There is little impact of bailouts on railroad employment, although the average wage paid by bailed-out railroads increased. Bailouts allowed loan recipients to reduce leverage, although we find no evidence that bond defaults declined. Interestingly, bailouts were associated with a reduced probability of a bond rating downgrade. Overall, there was a significant increase in bond prices following the announcement of a loan application or approval. Indeed, in the nine days surrounding news of an application, bond prices experienced an abnormal return of 2.2%. An approval is associated with a 1.8% abnormal return for bonds, but little effect on equity prices. Bailouts appear to have benefited existing employees and bondholders rather than aiding firm performance, such as railroad profits.

---

[1] RFC Final Report (1959), page 1.
[2] Over the entire period of the RFC's existence (1932-1957), the agency recovered 97.99% of the nominal value of the loans (see RFC Final Report, p. 163). We halt our examination in 1939 since the Great Depression is usually considered to have been over by the end of the 1930s, and World War Two began soon afterwards.
[3] In 1932 and 1933, RFC loans were extended at the same interest rate as Federal Reserve loans to member banks. See https://www.federalreservehistory.org/essays/reconstruction-finance-corporation (accessed April 26, 2022).

Second, we study the impact of railroad bailouts on the wider economy. Although New Deal railroad assistance was not explicitly aimed at the railroads' customers, we find that firms located in the same city or town as a bailed-out railroad experienced positive spillovers from news of a forthcoming railroad loan. Manufacturing firms whose operations had a high geographical overlap with the assisted railroad experienced a 0.9% abnormal return upon announcement of the bailout, compared to a 0.4% abnormal return for manufacturers with low levels of overlap with that railroad.

Railroads were a vital part of the U.S. economy in this era. The transportation sector employed over 3 million people in 1929 (6.9% of total employment), and generated $6.6 billion of national income (8.2% of total income), see Kuznets (1934). Railroads comprised roughly 10%, by number, of all NYSE stocks in CRSP in January 1929. By value, the railroad industry was about 13% of the market. There is extremely granular data on railroads, which allows us to know where railroads operated, the products they transported, and their employment levels. We also have annual financial statements and monthly revenue reports for all railroads.

Details of government railroad loans were quickly made public by the railroad regulator, the Interstate Commerce Commission (ICC), and widely reported in the media. It was, by contrast, impossible to observe the immediate impact of government loans in most sectors during the Great Depression. Loans to banks, farms, and industrial firms were largely kept secret, and financial claims on these firms were not usually traded in liquid financial markets. Anbil (2018) shows that banks that, unexpectedly, had their loans from the government revealed in August 1932 experienced drops in deposits of 18 to 25% compared to control banks. These banks were also 78% more likely to fail after their loans were publicly disclosed.

Government loans to railroads required approval by both the RFC and the ICC. The *New York Times* reported on May 21, 1932 that (p. 21), "although the corporation (RFC) cannot approve loans unless they are sanctioned by the (ICC) commission, it is not required to make the advances sanctioned." A loan approval required eight conditions to be certified by the RFC board of directors: (i) that the applicant could not obtain funds on reasonable terms from banks or the public, (ii) approval of the ICC, (iii) a maturity of less than three years, (iv) the borrower existed prior to 1932, (v) total government loans to the applicant were less than $100 million, (vi) no fees or commissions were paid by the applicant, (vii) the applicant consents to examinations by the ICC or other authorities, and (viii) collateral must meet the terms of the Finance Corporation act.[4]

In total, 191 railroad applications were successful, 32 applications were rejected, and fourteen applications were subsequently withdrawn. Railroads that obtained a government loan likely differed from railroads that did not receive government aid. Although we condition our results on the publicly available characteristics of railroads, it is likely that railroads also differed along unobservable dimensions. To address this issue, we take advantage of the political process that was inherent in RFC decision-making. That is, RFC directors were appointed by the President and confirmed by the U.S. Senate. Political considerations were important in the 'New Deal' decision-making process, see e.g., Wright (1974), Wallis (1987), and Fishback, Kantor, and Wallis (2003). We find that railroad bailouts were more likely to be granted to railroads that operated in the home states of RFC directors. Using the composition of the RFC board as an instrument for government assistance, we fail to find any beneficial effects on railroad employment or the likelihood of bond defaults.

Direct government aid to the real economy has been rarely attempted during a crisis, although direct aid was a big part of many governments' COVID-19 responses (see e.g., Cirera et al. (2021), Elenev,

---

[4] See *New York Times*, March 5, 1932, page 23.

Landvoigt, and Van Nieuwerburgh (2022), and Granja, Makridis, Yannelis, and Zwick (2020)).[5] As such, government aid to non-financial firms during a crisis has received little attention in the academic literature. Faccio, Masulis, and McConnell (2006) find that politically connected firms were more likely to be bailed out around the Asian financial crisis, especially if the national government had received an IMF or World Bank aid package. The authors conclude that bailed-out firms that were politically connected continued to underperform non-bailed-out firms in the same industry following the bailout, as measured by the return on assets (ROA). However, the ROA for non-connected firms improved relative to same-industry peers after a bailout. The study does not, therefore, fully determine whether real-sector bailouts are good public policy in a crisis. Goolsbee and Kruger (2015) argue that the bailouts of General Motors and Chrysler in 2008 helped to reduce the economic downturn in the U.S. They conclude, "The rescue has been more successful than almost anyone predicted at the time." Their study is necessarily restricted to two firms since the remaining Troubled Asset Relief Program (TARP) funds went to the financial sector. Berger and Roman (2017) investigate economic spillovers following TARP bailouts of U.S. banks. They find that in the counties in which banks received more TARP funds, there was better net job creation and faster wage growth — perhaps because TARP recipients passed on more generous loan terms to their customers (see Berger, Makaew, and Roman (2019)). Assistance first went to Wall Street before going to Main Street. While Berger and Roman study indirect assistance to the real sector, we study direct loans (at preferential interest rates) from the government to industry.

Non-financial firms experience a crisis differently to banks. There can be non 'runs' on a railroad's assets. U.S. railroads, for instance, often issued 50-year bonds to finance their operations, so there could be no 'run' on the railroad's debt unless their bonds were close to maturity (see Benmelech,

---

[5] Problems in the financial system during the Great Depression have received much attention (see, among others, Friedman and Schwartz (1963), Bernanke (1983), Calomiris and Mason (1997, 2003), and Benmelech, Frydman, and Papanikolaou (2019)). The consensus is that conditions in the financial and banking sector worsened the real effects of the Depression.

Frydman, and Papanikolaou (2019)). Railroads cannot quickly change their operations to 'gamble on resurrection' (see e.g., Hellmann, Murdock, and Stiglitz (2000) and Dewatripont and Tirole (2012)). Tracks, and other real assets, are fixed and costly to divert in the search for new business.

We discuss the economic environment that led to the creation of the RFC and the Corporation's structure in section 2. We describe our data and sources in section 3. We present our main results in section 4, with robustness checks in section 5. We conclude in section 6.

## 2. The Great Depression and the Reconstruction Finance Corporation

The Great Depression was an unprecedented period of economic and financial collapse worldwide. It struck the U.S. particularly severely with peak-to-trough industrial output falling 40% by late 1931 and GDP still 25% below trend six years after the recovery began (see Cole and Ohanian (2004) and Ohanian (2009)). There were several waves of banking crises in the early 1930s (see Bernanke (1983) and Friedman and Schwarz (1963)). In response to the weak economy and runs on troubled banks, President Hoover reluctantly created the Reconstruction Finance Corporation in January 1932, which was a component of what came to be known as the 'New Deal.' The RFC was initially permitted to loan to financial firms and railroads; loans were later permitted for farms, state and local government, infrastructure projects, and industrial firms. The RFC's initial capital stock came from a $500 million appropriation from the Treasury. While it obtained the bulk of its additional funding by issuing notes to the Secretary of the Treasury, a tiny part of its operations was provided for by direct borrowing from the public.

The *New York Times* reported on December 19, 1931, that Hoover believed that, "the plight of the American railroads is only temporary and that they will be able to work themselves out of the depression."[6] The United States had experienced severe railroad defaults during crises in 1873, 1884, and 1893, in which multiple large lines defaulted, resulting in significant drops in railroad employment (see Schiffman (2003), Giesecke, Longstaff, Schaefer, and Strebulaev (2011), and Cotter (2021)).

---

[6] *New York Times*, December 19, 1931, page 4.

Part of the rationale for providing aid to railroads was that many railroads were not capable of repaying their maturing bonds, and it would be exorbitantly expensive for them to obtain alternative funding from the banking sector. In late 1931, Daniel Willard testified in the Senate that railroads "cannot borrow money from banks at less than 8 or 9 percent interest" when most maturing bonds had coupon rates of around 4 percent.[7] The bond market for railroads virtually closed during the worst years of the depression, 1932 to 1934 (see Figure 1), and the government acted to fill that gap. Treasury Secretary Andrew Mellon saw the role of the RFC as to provide "a stimulating influence on the resumption of the normal flow of credit into channels of trade and commerce."[8] The Reconstruction Finance Corporation Act became law on January 22, 1932. The initial board of directors was appointed on February 2, and the first applications were received on February 5, 1932.

Mason and Schiffman (2004) calculate that, in 1931, 31% of railways' debt was held by insurance firms, 17% by banks, 4% by foundations and educational institutions, and 7% by other railroads, with the remainder held by "other" investors. Helping railroads avoid bond defaults would have been expected to reduce negative spillovers to the financial sector.

RFC loans to railroads were limited to three years duration, had to be 'adequately secured' by collateral, and were restricted to railroads that could not obtain funds on "reasonable terms". Collateral usually consisted of additional issuance of a railroad's bonds to the RFC, although occasionally direct liens on tracks or rolling stock was used. Most loans were for a three-year duration, and 83.5% of loans in our sample were rolled over at maturity. Railroads that were reorganizing under bankruptcy protection were also eligible for RFC loans. Overall, the RFC granted loans of, on average, $8.89 million.

---

[7] *New York Times*, December 23, 1931, page 16.
[8] *New York Times,* December 24, 1931, page 6.

Although disclosure of RFC loans to banks was very sporadic, the ICC had a policy of full and timely disclosure of railroad loans. All railroad loans were publicly disclosed at or near the time of loan application and approval. Loan information was sometimes, however, delayed slightly. The Baltimore and Ohio Railroad's loan application, for example, was kept secret for 10 days in August 1932. We show the distribution of railroad bailout loans over time in Figure 2. The biggest concentration of loans was in 1932 and early 1933, followed by another wave of new loans and rollovers in 1935 and again in 1938 and 1939. We depict the geographical distribution of loans by state in Figure 3. The large number of railroads that operated between New York and Chicago shows up as heavy approvals that went to railroads in the NorthEast and MidWest.

The initial board's ex-officio members were the Secretary of the Treasury, the Chairman of the Federal Reserve Board of Governors, and the Farm Loan Commissioner. Directorships were balanced by party affiliation, and care was taken that the directors came from different regions of the U.S. We read press reports, *Final Report on the Reconstruction Finance Corporation* (1959), and online biographies of the RFC directors to assign the directors' 'home' states. For example, the *New York Times* reported that two members of the initial RFC board would be "two Democrats from the Southwest, Harvey C. Couch of Arkansas and Jesse H. Jones of Texas."[9] We document the composition of the RFC board in Table 1, panel A. Most of the appointed RFC directors were businessmen, and four were former U.S. senators.

New Deal funding decisions are generally considered to have been at least partly politically motivated (see e.g., Wright (1974), Wallis (1987), and Fishback, Kantor, and Wallis (2003)). The RFC's decisions were similarly criticized. In April 1932, Representative La Guardia claimed, "everybody in the country knows a private wire from J. P. Morgan to the headquarters in Washington dictates the [RFC's] policy."[10] The RFC's initial president, former Vice President Charles G. Dawes, was heavily criticized by Senator Brookhart of Iowa for having loaned over $80 million to Dawes' own Chicago bank.[11] In our analysis, we demonstrate that RFC railroad loans were also partly determined by the geographical origins of the RFC board. We use the composition of the RFC board at the time loans were made as an instrument for loans.

---

[9] *New York Times*, January 26, 1932, page 1.

The Public Works Administration also made government loans to railroads from late 1933 until early 1936. PWA loans tended to be smaller than the RFC's disbursements, and they were often used for capital expenditure rather than to service the railroad's debt. In total, PWA loans only comprise around 10% of our sample by value, and 15% of our sample by number. Since money is fungible, we consider both RFC and PWA loans in our analysis.

### 3. Data

**3.1 Data sources**

We study the Class I railroads of the continental United States. Class I railroads owned over 90% of the nation's tracks by length, they employed roughly 98% of railroad employees (3.4% of the United States' total labor force), and they carried over 99% of the revenue-ton-miles of all U.S. railroads in 1929. We collect balance sheet, profit and loss, track network, and employment data for these railroads from the annual reports of the ICC, *Statistics of Railways in the United States*.

We compile annual statistics for each railroad's freight revenue sources (i.e., agricultural, animal, mining, forestry, merchandise, or manufacturing items) and monthly revenue from *Moody's Manual of Investments – Railroad Securities*. Data on freight revenue is important because some railroads concentrated on transporting a narrow range of products. As a result of the railroads' varied exposure to product markets, the Great Depression could affect railroads unevenly. Therefore, we use the ICC's classification to place railroads into eight geographical regions.

---

[10] *New York Times,* April 7, 1932, page 2.
[11] See *New York Times,* September 16, 1932, page 2.

If a railroad had publicly traded equity, we gather stock prices from the Center for Research in Security Prices (CRSP). Many railroads did not have publicly traded equity, usually because they were fully owned by another railway or a related industrial firm. We compile price data on the most liquid bond per railroad.[12] We obtain bond prices from the section 'Bond Sales on the New York Stock Exchange' in the *New York Times*. We classify a railroad as being in default if it had failed to meet a coupon or principal repayment or if it in any way changed the terms of the issue.[13] Data on bond ratings, coupons, amounts outstanding, and maturity comes from *Moody's Manual of Investments – Railroad Securities*. We use Moody's index of daily corporate bond prices, reported in the *Commercial and Financial Chronicle*, as a proxy for the market return on railroad bonds.

We combine the track network of each railroad with two city-level data sources. First, we hand-collect data on factories operated by NYSE-listed manufacturing firms from *Moody's Manual of Investments – Industrial Securities*. In total, there were 471 manufacturing firms that have data on factories reported in *Moody's*. Second, we obtain city-level building permit data from Cortes and Weidenmeir (2019). The value of these permits is based on the costs of new commercial and residential buildings for 215 cities across the U.S., taken from *Dun & Bradstreet's Review*. We collect national bank capital in default from *Annual Report of the Comptroller of the Currency*. National bank capital per state comes from Flood (1998).

---

[12] There are 46 railroads with listed equity and 72 railroads with bonds.
[13] For example, extending the maturity of the bond, reducing the coupon rate, or exchanging the initial bond for another security.

## 3.2 Bailouts

We identify railroad bailouts from the Reconstruction Finance Corporation's records at the National Archives in Washington.[14] For each railroad, we observe their application, approval, and refusal documents and the dates on which each document was sent. Loan decisions were made quickly, usually taking a couple of weeks to a month or two. The median (mean) value of *Decision time* is 27 (45) days.

To identify the public announcement date of railroad bailouts, we search the *New York Times* for the phrases "Reconstruction Finance Corporation" or "Public Works Administration" from January 1932 until December 1939. We identify loan applications, approvals, rejections, and details thereof in the archives. 89.5% of all applications, approvals, and refusals from the archives also appear in the *New York Times*.

Occasionally a railroad would revise the size of their loan request while the application was under consideration. For example, the RFC might agree to loan a railroad a smaller amount than initially requested and would invite the railroad to modify its application. Alternatively, a railroad might have been successful in obtaining some funds from other sources and reduce their loan amount. For every loan we calculate the difference between the application and approval sizes (*Loan size difference*).

## 4 Results

### 4.1 RFC Board composition

In Table 1, panel A, we present data on RFC directors. The composition was supposed to be balanced by party affiliation and geographically diverse. However, it is possible that the appointment of RFC directors was partly determined by economic conditions in the home state of the director or even by financial conditions in the railway sector in their home states. In Panel B, we document that larger states were more likely to have RFC directors, and we find evidence that directors were less likely to be appointed if there was already a director from the same state. Our results show that the appointment of directors is not robustly associated with economic conditions in the directors' home states. Therefore, concerns are alleviated that causality runs from the poor economic conditions of railroads to the appointment of RFC directors and thence to more railroad bailouts.

---

[14] Records RG 234.

**4.2 Summary statistics**

In Table 2, we present our summary statistics on railroads. We divide railroads into those that were "bailed-out"—which we define as having received at least one loan from the RFC or the PWA between 1932 and 1939—and those that were not bailed-out. In Panel A, we show that there are large differences between the bailout recipients and others. Bailed-out railroads had less cash to total assets (a mean of 1.5% of assets vs. 2.3% for non-bailed-out railroads), were slightly more levered (mean long-term debt to total assets of 42.5% vs. 40.7%), were less profitable (a mean net income to total assets of 0.9% of total assets vs. 1.3%), and had more volatile operations (monthly volatility of 12.3% vs. 9.2%). Bailed-out railroads were much larger (mean total assets of $195.1 million vs. $29.2 million), employed more people (a median of 9,145 vs. 1,214), had higher average wages ($1,634 vs. $1,568), and had more connections to the RFC board. On average, bailed-out railroads operated in 0.8 states with an RFC director vs. 0.4 for non-bailed-out railroads.[15]

In Panel B, we subdivide the bailed-out railroads into two groups: those that received a single loan from the RFC or PWA and those that received multiple bailouts between 1932 and 1939. The railroads that received multiple bailouts tended to have less cash (mean 1.4% of total assets vs. 1.9%) and were better connected to the RFC (connections in 0.9 states vs. 0.4 states). The companies that obtained multiple bailouts tended to be larger (mean size of $247.0 million vs. $65.1 million), employed more people (a mean of 10,076 people vs. 2,609 people), and focused more on passengers (12.5% vs. 7.6% of total revenue at the mean).

---

[15] The statistics in Panel A combine observations before and after the loans. We also examine *ex-ante* differences between loan recipients. The differences between bailed-out railroads and non-bailed-out railroads in 1929 mirror those in the full sample (results available upon request).

**4.3 Bailout recipients**

We use a two-step model to investigate which railroads received the government bailouts, in the spirit of Vossmeyer (2016). In the first step, we run an OLS model of railroads' applications on their lagged characteristics. *Application* equals one if the firm applied for at least one bailout in that year, and zero otherwise. We include year, bond rating, and region-year fixed effects to control for unobserved, time-varying shocks to economic conditions and RFC board processes.

In column 1, we highlight that railroads were more likely to apply for a bailout when they had more connections to the RFC. An increase in the number of connections of one leads to an increase in the probability of applying by 0.029 (significant at the 1% confidence level). Given that the average probability of applying equals 0.067, this effect is economically large. Railroad characteristics, such as cash holdings, leverage, and profitability do not change the likelihood of a railroad applying for a bailout, once we condition on a railroad's bond rating. In column 2, we allow for unobserved, time-varying shocks at the regional level. Allowing for these shocks does not change the coefficient estimates.

In the second step, we study which characteristics correlated with a loan approval, among railroads that applied for (at least) one loan. The dependent variable, *Approval*, equals one if a railroad received at least one bailout in a year in which it applied at least once. In columns 3 and 4, we find that successful applicants were, on average, larger, more levered, focused more on passengers, and had more bonds that were close to maturity.[16] Applicants that had received more loans in the past were less likely to be approved. When we take the unobserved, time-varying regional effects into account (column 4), we see little change.

In columns 5 and 6, we examine the size of loan approvals (scaled by total assets). We see that more levered railroads tended to receive larger loans, and applicants who had received more prior loans from the government tended to receive smaller loans. Railroads whose leverage was one standard deviation higher (15.9 percentage points of total assets) received a bailout that was, on average, 38.92% higher.

### 4.3.1 Market reactions to bailouts

We examine the reactions of a railroad's stock and bond prices to *New York Times'* reports of bailout applications and approvals. In Table 4, we compute abnormal returns (AR) and cumulative abnormal returns (CAR) on railroad debt and equity. We compute abnormal returns as the return on the stock or the bond less the CRSP market return or Moody's corporate bond index return, respectively.[17]

We find statistically significant abnormal returns of 0.9% for bonds on the day a loan application was announced and a statistically insignificant abnormal return of 0.3% on the day an approval was announced (see Panel A). There is no statistically significant AR on refusal announcement dates, although there is a significant abnormal return the next day of -5.1%. In the window around the news release (t-4 to t+4), we find CARs of 2.2% (applications), 1.8% (approvals), and -6.6% (refusals), although the refusal return is statistically insignificant.

---

[16] We include a railroad's debt maturing in the depths of the depression following the papers of Benmelech, Frydman, and Papanikolaou (2019) and Duchin, Ozbas, and Sensoy (2010).
[17] We only hand-collect bond prices in a narrow window around RFC announcements. Therefore, we are unable to estimate a market model for railroad bonds. To maintain consistency in our measurement of abnormal returns between bonds and stocks, we compute abnormal returns for railroad stocks in the same manner. Effectively, we assume that alpha equals zero and beta equals one in the market model.

Since many railroads applied for (and were granted) multiple loans, we investigate the differences between the initial loan and subsequent loans. Substantially more private information is likely to have been conveyed to the market by a firm's initial revelation that it desired federal government financial assistance. An application announcement for the first bailout is associated with a 4.1% bond CAR from t-4 to t+4, and a 2.1% AR on day zero (see Panel B). An approval announcement for the first bailout has a 1.2% bond AR on day zero (with a 5.9% CAR from t-4 to t+4), all statistically significant. Subsequent bailouts are reflected in more subdued bond responses. A second or subsequent application has a 0.4% bond AR on announcement day (1.4% from t-4 to t+4), both statistically significant. A second or subsequent approval has no movement on day zero, but a 0.6% AR (significant) on day one and a statistically insignificant 0.9% CAR from t-4 to t+4.

In contrast to the response of bond prices, there is little statistically significant movement of stock prices in response to bailout news events (see Panels A and B). News of subsequent RFC approvals resulted in a 1.5% AR on the announcement date and a 2.3% CAR (both statistically significant) from t-4 to t+4.

### 4.3.2 Determinants of announcement returns

We examine the association between a railroad's characteristics at the time of the bailout and its announcement returns. In Table 5, we regress the CAR of the railroads' bonds and equity (from t-4 to t+4) on firm and bailout variables. We find few railroad characteristics that are robustly associated with security returns. Most characteristics are insignificant and/or change signs depending on whether we examine applications versus approvals or bonds versus shares. Railroads with more leverage experience substantially worse returns on their equity upon announcements of loan applications, perhaps because a loan application indicated the railroad would struggle to service its debt, and, therefore, that equity was next to worthless. A one standard deviation increase in leverage corresponds to a 9.60% more negative CAR. Railroads with more employees tended to have lower announcement returns, perhaps because market participants expected government pressure on the railroad to maintain employment.[18] A one standard deviation increase in employment is associated with a 2.03% more negative CAR.

## 5 Effectiveness of government bailouts

We now examine if railroad bailouts achieved the RFC's twin objectives. We first investigate if the bailouts helped the railroads avoid defaulting on their bonds. Second, we look at how the bailouts affected railroad employment, wages, and operating performance. Finally, we search for evidence that these bailouts provided positive spillovers for firms that relied on the railroads to provide transport services.

**5.1 Credit structure**

Did the RFC protect the "credit structure" of the financial system? All else equal, an RFC loan should have made a railroad less likely to default on its debt. Jones (1951) claims that RFC funding reduced railroad defaults by half, whereas Schiffman (2003) and Mason and Schiffman (2004) claim that bailouts at least delayed defaults.[19] However, defaulting on debt is partly a choice, and Mason and Schiffman argue that bankruptcy "brought relief from high fixed charges that were often a principal cause of financial distress" (p. 61). In Figure 4, we plot Kaplan-Meier (1958) graphs with cumulative probabilities of failure for the bailed-out vs. non-bailed-out railroads. We observe that railroads that received a bailout are associated with a higher hazard rate of bond defaults, and that this difference increases with time from the bailout. In Panel B, we show that the higher default rate for bailed-out railroads survives the inclusion of control variables. The granting of a below-market-rate loan, *ceteris paribus*, is a good event. Hence, higher default rates for bailed-out railroads suggest that unobservable factors are likely influencing both a railroad's performance and the bailout decision.

---

[18] An alternative explanation is that investors may have perceived that railroads that received bailouts would be more generous with their employees' compensation.
[19] As Jones was RFC chairman for much of the 1930s, his claims of RFC loan effectiveness should be taken with a grain of salt.

We assess the effects of bailouts on bond defaults in Table 6. In column 1, we run a probit model of defaults in which the dependent variable equals one if a railroad defaulted on its bonds in that year. We examine if loan *Approval* in the previous year is associated with the railroad defaulting upon its debt. We attempt to capture railroad unobservables by including bond rating fixed effects from Moody's. In this era, Moody's only released ratings once per year in its annual investors' compendium. Most railroads had multiple bonds and bond ratings so we use the rating of the bond closest to maturity. Overall, we show that lower net income, lower cash to assets, and younger railroads are associated with a higher likelihood of default. Government loan *Approval* does not have a statistically significant relation with defaults.

In column 4, we again run a probit model of defaults, but the dependent variable now equals one if a railroad defaulted on its bond in that year or the following two years. This offers a longer-run investigation of how a government loan *Approval* is associated with a railroad's likelihood of default. We find that *Approval* has a significantly positive correlation with the probability of railroad default (at the 5% confidence level). Indeed, getting an RFC or PWA bailout is associated with a default rate of 6.39%, all other characteristics at sample means, relative to a non-bailed-out railroad's default rate of 1.51%. This finding is in line with the Kaplan and Meier graphs in terms of magnitude.[20] However, railroad bailouts are unlikely to be awarded at random, and a selection effect is likely to be present. Hence, we turn to an instrumental variable approach to determine if bailouts have a causal effect on railroad defaults.

### 5.1.1 Instrumenting for bailouts

Our concern in determining if bailouts aided railroads is that there are likely to be variables omitted from our econometric specification that partly affected a railroad's default behavior. Railroad management and the RFC board had access to non-public information. The RFC required all railroads that applied for loans to disclose the "possibility of securing a loan from other sources" (item 3 of the application). For example, the Chicago and NorthWestern Railroad, in applying for $11,127,700 on January 16, 1933, reported to the RFC that "for many years Applicant ordinarily has been financed through Kuhn, Loeb and Company, Investment Brokers of New York City … on different occasions during the latter part of 1932 Applicant discussed … the possibility of securing a loan … but Kuhn, Loeb and Company declined to commit itself to any future loans." Railroads that had tried, but failed, to obtain bank or Wall Street assistance to raise additional funds would be more likely to default than its observable characteristics would suggest. In such a situation, the error from the regression of bond defaults on bailouts is likely to be correlated with the independent variable *Approval*. Therefore, the coefficient estimates on *Approval*, which measure the effectiveness of government aid, will be biased.

---

[20] This finding is robust to changes in the regression specification, such as OLS fixed-effects models.

We would like to use an instrumental variable that is correlated with a railroad receiving a government bailout but only affects a railroad's financial performance via the granting of RFC loans. We take advantage of the prior literature (see e.g., Wright (1974), Wallis (1987), and Fishback, Kantor, and Wallis (2003)) that claims New Deal grants were influenced by politics. Fishback (2017), for example, concludes, "nearly every study finds that political considerations were important to the Roosevelt administration." There are, however, a few investigations of New Deal funding, such as Mason (2003), that find little political influence on the process. We use the composition of the RFC board as our instrumental variable. Specifically, we use the number of states a railroad passed through that were the home states of RFC directors in that particular year. We call this the number of a railroad's *Connections* to the RFC. For example, on February 5, 1932, the Chicago and Eastern Illinois railroad applied for an RFC loan for $3.629 million. This railroad passed through Illinois, Indiana, and Missouri. H. Paul Bestor (Missouri) and Charles G. Dawes (Illinois) sat on the board of the RFC at the time of the application. Therefore, our instrument takes a value of two.

In our first stage regression (Table 6, columns 2 and 5), we regress *Approval* on a railroad's lagged characteristics and our instrument, *Connections*. We see that *Approval* is positively and statistically significantly related to a railroad's *Connections*, even with region, year, and bond-rating fixed effects. The F-statistic in the first stage regression is 75.791, which indicates that we have a strong instrument.

To have a valid instrument, we also require that the exclusion restriction is satisfied. The exclusion restriction requires that *Connections* are uncorrelated with the error term, the unobservable part of a railroad's financial position that partly determines default behavior. There was no realistic possibility that a railroad that was doing poorly, based on unobservable factors, could increase its number of *Connections* by altering its operations. Total track mileage in the U.S. declined from 1930 onwards, and it would be expensive and take years of construction for an existing railroad to begin operations in the home state of an RFC director.[21]

It also would invalidate our instrument if railroads that were in worse financial shape than their observable characteristics would suggest were able to influence the president to alter the RFC board's composition, such that a new director was appointed from a state in which the railroad operated. RFC directors were responsible for approving all loans that the Corporation made, railroad and non-railroad alike. Total railroad loans comprised a fraction of the RFC's disbursements, and loans to an individual railroad were a tiny percentage of total RFC expenditure. There were only five to seven directors at any one time, and the composition was balanced by political affiliation and by the need to have directors come from different parts of the country. Given these constraints on the composition of the RFC board, we believe it is extremely unlikely that certain railroads could have increased their *Connections* by lobbying. Therefore, we use *Connections* as our instrument for RFC bailouts.

In columns 3 and 6 of Table 6, we replace *Approval* with its predicted level from our first-stage regression. We see that bailouts do not have a statistically significant relation with a railroad's defaults. Government aid did not seem to aid railroads avoiding default. Other coefficients show that railroads that were more profitable, had more cash, and were older were less likely to default.

---

[21] See ICC *53rd Annual Report on the Statistics of Railways in the United States* (1939), Table 1-A.

### 5.1.2 Bond ratings

Bailed-out railroads may have been viewed as "too big to fail" or perhaps investors anticipated that a bailout indicated that the government would share the financial losses with bondholders.[22] In this case, the bond market may have perceived the railroad's debt as being safer, despite our evidence in Table 6 showing that bailouts did not help a railroad to avoid default. To investigate perceptions of default, we examine a railroad's bond ratings.

In Table 7 we study the determinants of ratings changes. We examine the effect of receiving a bailout (*Approval*) last year, and condition on last year's observable firm characteristics. A bailout reduced the probability of being downgraded in the current year from 0.258 to 0.178 if the railroad was bailed out in the previous year (column 1). The probability to receive multiple downgrades, from year *t* to *t+2*, was reduced from 0.350 to 0.186 if the firm was bailed out in year *t-1* (column 4). Railroads with more cash and lower employment were less likely to be downgraded.

Since concerns about selection effects remain, we again run instrumental variable probit regressions with *Connections* as our instrument. The first-stage regressions appear in columns 2 and 5, and the second-stage results are in columns 3 and 6. The IV results confirm the probit results. Bailouts in the previous year drive a decrease in the likelihood of getting one downgrade of 52.7% (column 3) or a decrease in the likelihood of getting multiple downgrades of 92.2% in the subsequent three years (column 6). Railroads with more employees, less cash-to-assets, and younger railroads were more likely to be downgraded. This result suggests that Moody's perceived increased railroad employment during the Great Depression to be in conflict with protecting bondholders' interests. Overall, our results show that government bailouts did not protect railroads against default, although they did alleviate bond ratings downgrades.[23]

---

[22] The RFC rarely insisted on holding super-senior claims on a railroad's debt. It would often agree to hold mid-tier claims on a railroad's assets.

## 5.2 Operating performance

### 5.2.1 Difference-in-differences

We now investigate if the RFC succeeded in their objective to "stimulate employment." As such, we determine the ability of government bailout recipients to improve their economic performance, including employment numbers. We first conduct a difference-in-differences approach on RFC loan recipients' leverage, employment, the average wage, and profitability. Since we have variations in treatment timing, we use the technique of Callaway and Sant'Anna (2021). The treatment group is railroads that received at least one bailout in the sample period, the control group are railroads that never received a bailout.[24]

In Panel A, we present average treatment effects (ATT). We document a significant negative impact of a bailout on leverage (column 1). Indeed, the estimated treatment effect on leverage is a reduction of 2.6 percentage points for the bailed-out railroad. The starting leverage for the average railroad was a little over 40 percent of total assets (see Table 2), so this coefficient is economically meaningful. This result is also in line with the theoretical contribution of Elenev, Landvoigt, and Van Nieuwerburgh (2022), who study the effect of COVID-19 bailouts on the real economy.

We find positive but statistically insignificant effects of a bailout on employment (Column 2). There are statistically significant increases in the average wage of around 4.5% (Column 3). Our results of weak employment and generous wages align with the findings of Ohanian (2009) and Cole and Ohanian (2004) that New Deal policies deepened the Great Depression.[25] Railroads appear to have used (some) government funds to inflate the average wages they pay their employees. Finally, we find statistically insignificant effects of a bailout on profitability (Column 4).

---

[23] In Table 6, Moody's ratings add information to understand bond default dynamics, even after controlling for the railroad characteristics. However, bond ratings mostly help to explain the default behavior of non-investment grade bonds. The only bond rating fixed effects which are significantly different to zero are the lowest rated (e.g., C, Ca, and Caa). Most government bailouts went to railroads with investment-grade bonds. Government bailouts helped such bonds to preserve their (high) credit rating (Table 7) and since investment-grade bonds are unlikely to default, a bailout did not greatly change their default risk (Table 6).

[24] Using a chi-squared test we verify that the parallel trends assumption is never violated.

In Panel B, we present estimates of the treatment effect by year. We find that leverage decreases in years t+1 through t+4, although it is barely affected in the year of the bailout. We find employment of bailed-out railroads is statistically significantly higher than the control group in years t+1 and t+2 (by 4.5% and 6.9% respectively), although by year 3 and 4 the point estimate is lower and no longer statistically significant. Bailout recipients' average wages were higher in the year of the bailout and remained higher, by a little under 5%, in the following years. There are no statistically significant impacts on profitability after the treatment.

In Panel C, we run a placebo test in which we counterfactually assume that RFC bailouts took place in 1929, while still comparing bailed-out and non-bailed-out railroads (as in Berger and Roman (2017)). We use 1927 and 1928 as the "pre-RFC period" and 1930 to 1932 as the "post-RFC period". We apply the doubly-robust difference-in-differences approach of Sant'Anna and Zhao (2020). In Panel C, we fail to find significant results for leverage, employment, average salaries, or profitability.

### 5.2.2 Instrumental variables

Although the difference-in-differences approach should give us a good idea of the impact of a railroad bailout, there remain concerns that the comparison group of non-bailed-out railroads does not provide an accurate counterfactual for the bailed-out carriers. In Table 9, we again make use of the board composition of the RFC and our measure, *Connections*, as an instrument for railroad bailouts. In the second stage, we regress railroad leverage, employment, wages, and profitability on the fitted level of bailouts after conditioning on railroad characteristics.

---

[25] Schiffman (2003) finds that railroads that defaulted increased their employment following the default.

We find that a bailout causes a 10.2 percentage point decrease in leverage (column 2). We find no statistically significant impact of bailouts on employment (column 3), or profitability (column 5). However, there was an increase of 15.4 percentage points in the average wage (column 4). Therefore, we conclude that the RFC failed in their second objective, which was to promote railroad employment. All regressions use firm, year, and region fixed effects and condition on lagged characteristics. Railroads appear to have used bailouts to reduce their leverage, and increase wages, with little beneficial impact on employment.

**5.3 Economic spillovers**

Bailouts do not seem to have provided any direct benefits for the recipients—save a jump in the value of their debt. They may, however, have provided spillover benefits for the regions in which they operated. For example, railroads may have been able to keep operating routes that would otherwise have been closed, or the carriers may have conducted a more frequent schedule for local businesses than if government support had not been made available.

**5.3.1   Building permits**

We examine if there were positive economic spillovers that flowed from the bailouts of railways that passed through a city. We create an explanatory variable, *City RFC Approvals*, which equals the fraction of all railroads that passed through a city that received an RFC or PWA loan in the previous year. We again use our instrumental variable, *Connections*, which is measured at the state level, as an instrument. Our focus is on spillovers to one of the few measures of local economic conditions available in this era, building permits (see Cortes and Weidenmeir (2019)).

We regress the natural logarithm of city building approvals per capita in a year on fitted *City RFC Approvals*.[26] In Table 10, we see that RFC board connections are very strong instruments for city-level loan approvals. We find a negative relationship between railroad city-level loan approvals and new building approvals (columns 2 and 4). Once we add both year- and city-fixed effects, however, the estimated impact of *City RFC Approvals* on building approvals becomes statistically insignificant and close to zero in magnitude (column 6).

### 5.3.2 Manufacturing firms

In Table 11, we study if news of a railroad's bailout affected other railroads and manufacturing firms listed on the NYSE.[27] We calculate the abnormal returns and cumulative abnormal returns of other firms' equity. We focus on the more interesting cross-sectional evidence: which manufacturing firms and which railroads benefited most from news of one railroad's bailout?

We cross-sectionally split firms into two dimensions. First, did the other railroad overlap at all with the bailed-out railroad, meaning did both railroads service at least one common city (*Yes*) or not (*No*)? Second, was the level of overlap (the fraction of the bailed-out railroad's cities also serviced by the other railroad) above the sample mean (*High*) or was the overlap positive but below the sample mean (*Low*). We construct similar overlap measures for manufacturing firms, but we consider the joint presence of manufacturing establishments (from *Moody's Manual of Investments – Industrial Securities*) and railroad tracks.

In Panel A we see that there was no statistical difference between the the mean CARs of railroads that overlapped with the bailed-out railroad and non-overlapping railroads upon news of an RFC application. Nor was there a statistically significant difference between high overlap railroads and low overlap railroads. In Panel B, we observe slightly smaller, -0.5% (*Yes* less *No*) to -0.6% (*High* less *Low*), and statistically significant differences in CARs for other railroads upon RFC approvals. We interpret this result to mean that competing railroads (i.e., those with some overlap with the bailout recipient) suffered slightly from a bailout announcement, relative to railroads that had little or no overlap. As this is a cross-sectional test, we are conditioning on any economy-wide railroad shocks, such as changes in government railroad policy, input costs, or demand changes.

---

[26] Our thanks to Gustavo Cortes for sharing his data on building approvals.
[27] We exclude all stocks that have a zero return on all days of the event study.

We compare manufacturing firms that were co-located in the same city as the bailed-out railroad (an overlap of *Yes* or *High*) to manufacturing firms that were not located in cities through which the bailed-out railroad ran (*No* or *Low*). We see that a railroad bailout benefited co-located manufacturing firms relative to manufacturers that were not located near the bailout recipient's tracks. Co-located manufacturing firms outperformed others by 0.1% (*Yes* vs. *No*) at the time of the application and 0.2% at the time of the approval. If we compare *High* vs. *Low* manufacturing firms, we see that high-overlap manufacturers outperformed by 0.6% at the time of application (Panel A) and 0.3% at the time of approval (Panel B). Taken together, there is evidence that railroad bailouts provided positive economic spillovers to the real economy, even if the railroads themselves showed little direct benefit from the bailout.

## 6 Conclusion

The RFC distributed much of the U.S. government's New Deal assistance to the economy as it struggled with the Great Depression. Around 10% of the RFC's loans were given to private firms in the railroad sector, combined with a limited amount stemming from the Public Works Administration. In our study, we ask if such bailouts aided railroads to avoid debt defaults and stimulate employment, which were the RFC's twin objectives.

First, we find scant evidence that government assistance was beneficial for railroad employment, although it led to an increase in wages for existing employees. Second, there is no evidence that government aid prevent bond defaults. However, we show that bailouts helped railroads to reduce leverage and assisted them to avoid ratings downgrades. This is also reflected by a jump in the bond prices on the days around loan announcements or approvals. As such, we argue that bailouts benefited employees and bondholders rather than aiding firm performance (e.g., profitability).

Non-bailed-out railroads that competed with the recipient seem to have suffered some harm from the bailout, presumably because one of their competitors was supported financially. We find evidence that government railroad support was beneficial for the manufacturing firms that were co-located near the railroad's tracks. As such, although RFC and PWA assistance proved of little benefit to the railroad itself, there were positive economic spillovers to manufacturers from this New Deal program.

**Figure 1: New railroad bond issues**

Number of new bond issues by all class I railroads between 1927 and 1939.

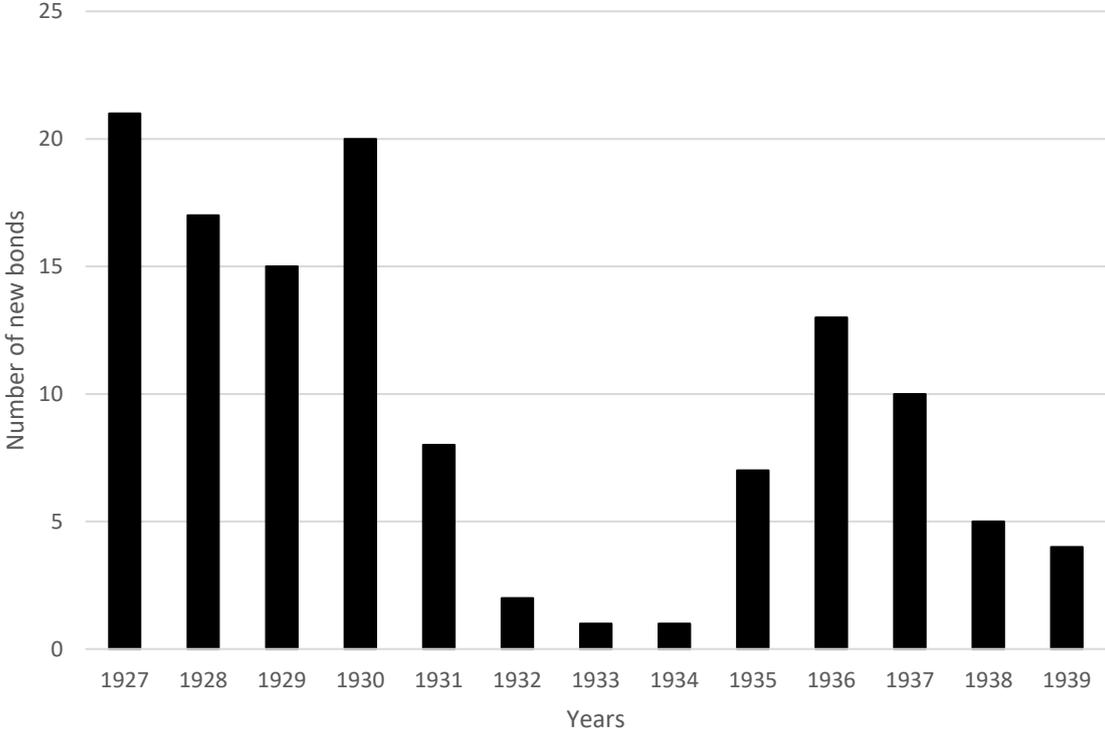

**Figure 2: RFC railroad loans ($ million, including rollovers)**

Value of approved bailouts for all class I railroads between 1932 and 1939 on a quarterly basis.

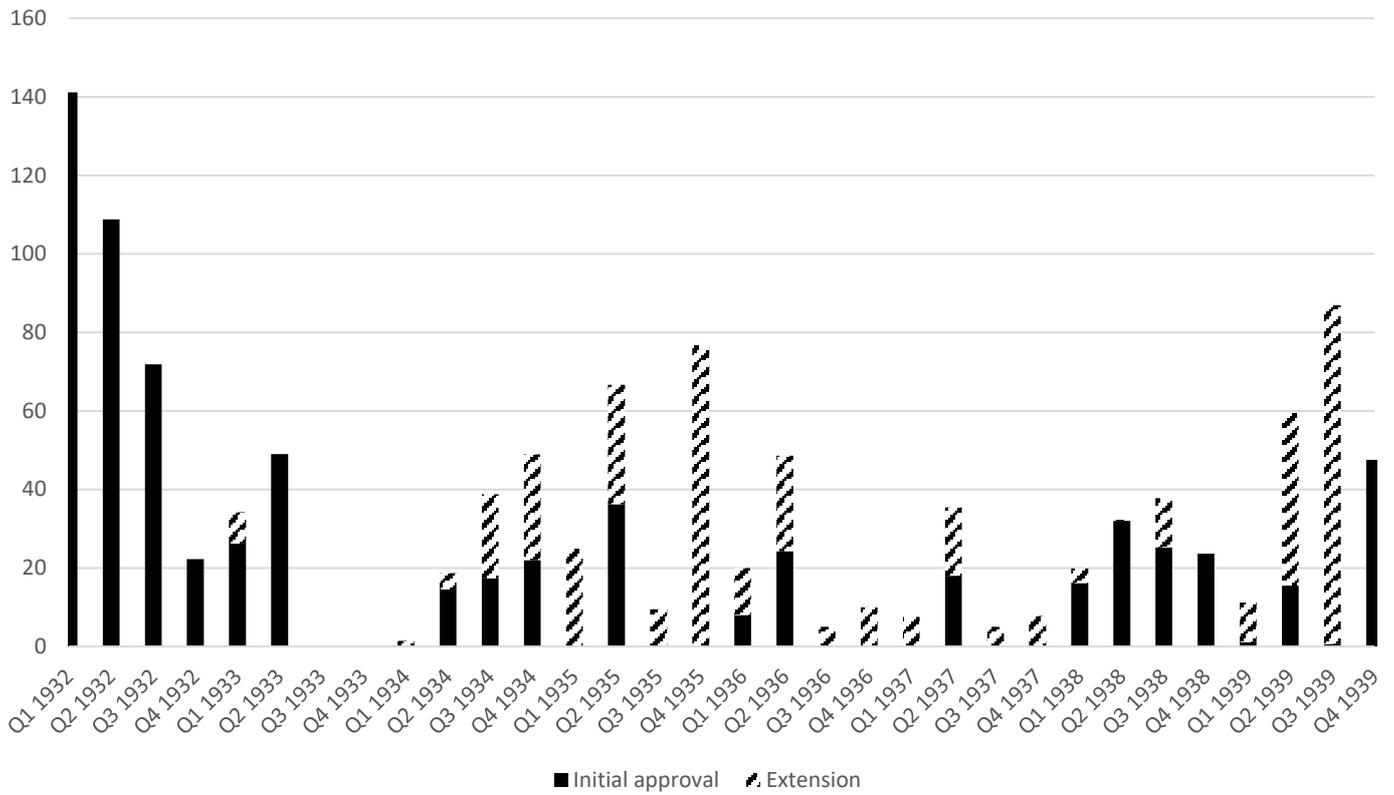

**Figure 3: Number of loan approvals**

The number of RFC or PWA loans to railroads that operated in each state. For railroads that operated in more than one state, we count each state in which that railroad operated as having received a loan.

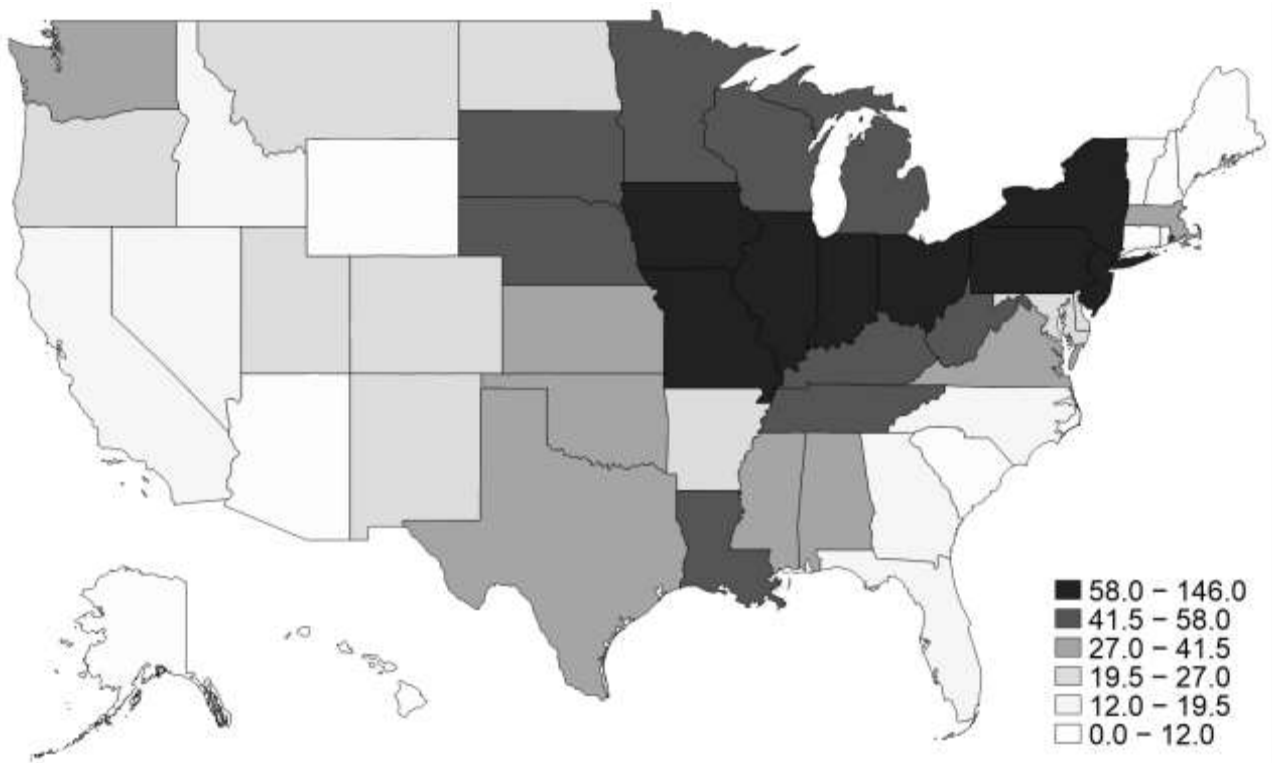

**Figure 4: Kaplan-Meier failure graphs**

We show the hazard rates of bond defaults in the years after a bailout. Panel A shows the hazard rates for bailed-out vs. non-bailed-out railroads. Panel B shows hazard rates after controlling for lagged railroad characteristics, such as (log) total assets, net income to total assets, cash to total assets, leverage, (log) employment, (log) firm age, bonds due between 1930 and 1934 to total assets, passenger revenue to total revenue, and the freight composition.

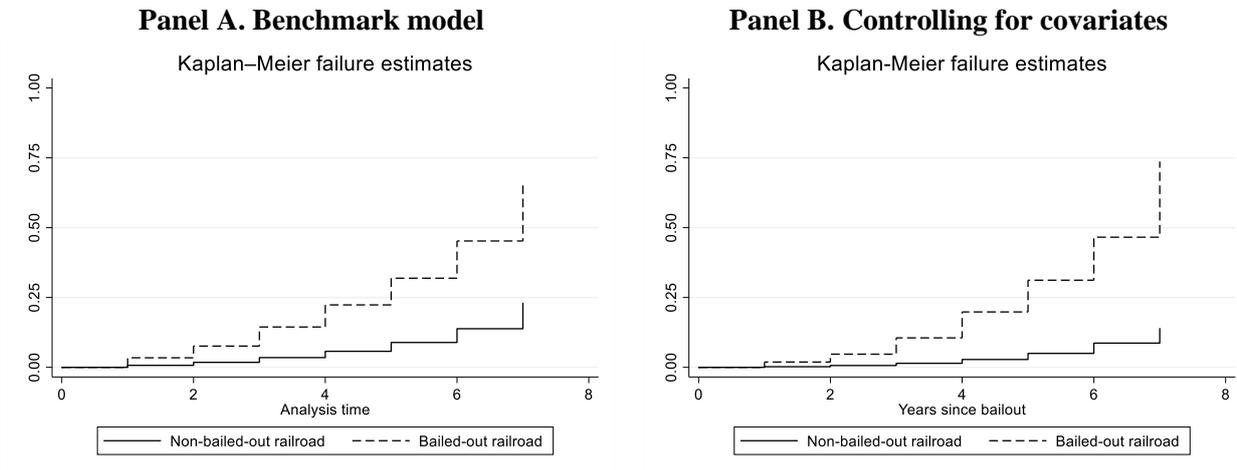

## Table 1: RFC board composition

Panel A reports the RFC board composition between 1932 and 1939. *Party* refers to the respective RFC director's political party, where *Dem.* refers to Democrat and *Rep.* refers to Republican. *State* refers to the respective director's home state (abbreviated). *Start* and *End* refers to the director's start and end time on the board, respectively. *Comments* gives insights into their background. In Panel B, the dependent variable equals one if an RFC director from state *y* was appointed to the board in year *t*, zero otherwise. Column 1 includes all state-years; column 2 excludes appointments to the first board; column 3 examines RFC directors with previous political experience; column 4 examines RFC directors from the private sector. *Bank Capital in Default* is the ratio of national bank capital in default to total national bank capital in state *y*. Log(*Size per Railroad*) is the logarithm of the total assets of railroads active in state *y* divided by the number of railroads active in state *y*. *Number of Railroads* is the number of railroads active in state *y*. *Railroad Bailout Weight* is the ratio of the total assets of bailed-out railroads in year *t* active in state *y* to the total assets of railroads active in state *y*. Log(*Building permits per capita*) is the log of building permits divided by the state population in state *y*. *Banks per capita* is the number of national banks in state *y* divided by state population. *Number of RFC directors* is the number of existing RFC directors from state *y*. All regressions use year and state fixed effects. We cluster standard errors at the state level. *, **, and *** denote significance at the 10%, 5% and 1% levels, respectively.

| Panel A: RFC Board composition | | | | | |
|---|---|---|---|---|---|
| Name | Party | State | Start | End | Comments |
| H. Paul Bestor | Rep. | MO | Feb-32 | Jul-32 | ex officio as Farm Loan Commissioner |
| Eugene Meyer | Rep. | NY | Feb-32 | Jul-32 | ex officio as governor of Fed. Reserve |
| Andrew W. Mellon | Rep. | PA | Feb-32 | Feb-32 | ex officio as Treasury Secretary |
| Ogden L. Mills | Rep. | NY | Feb-32 | Mar-33 | ex officio as Treasury (Under) Secretary |
| William Woodin | Dem. | NY | Mar-33 | Dec-33 | ex officio as Treasury Secretary |
| Arthur H. Ballantine | Rep. | - | Feb-32 | May-33 | ex officio as Treasury Under Secretary |
| Dean H. Acheson | Dem. | - | May-33 | Nov-33 | ex officio as Treasury Under Secretary |
| Henry Morgenthau (Jr.) | Dem. | NY | Nov-33 | Feb-38 | ex officio as Treasury (Under) Secretary |
| Thomas J. Coolidge (III) | Dem. | MA | May-34 | Feb-36 | ex officio as Treasury (Under) Secretary |
| Roswell Maginn | Dem. | IL | Jan-37 | Feb-38 | ex officio as Treasury (Under) Secretary |
| Harvey C. Couch | Dem. | AR | Feb-32 | Aug-34 | Arkansas businessman (electricity, railways) |
| Charles G. Dawes | Rep. | IL | Feb-32 | Jun-32 | Former Vice President and Chicago banker |
| Jesse H. Jones | Dem. | TX | Feb-32 | Jul-39 | Texas businessman (lumber, real estate, banking) |
| Wilson McCarthy | Dem. | UT | Feb-32 | Sep-33 | Utah state senator and district attorney |
| Gardner Cowles (Sen.) | Rep. | IA | Jul-32 | Apr-33 | Des Moines newspaper proprietor |
| Charles A. Miller | Rep. | NY | Aug-32 | Mar-33 | Utica banker |
| Atlee Pomerene | Dem. | OH | Aug-32 | Mar-33 | Ohio lawyer and former U.S. senator |
| Carroll B. Merriam | Rep. | KS | Jun-33 | Dec-41 | Topeka finance industry |
| John J. Blaine | Rep. | WI | Jun-33 | Apr-34 | Lawyer and businessman, former U.S. senator |
| Frederic H. Taber | Rep. | MA | Jun-33 | Jan-38 | New Bedford lawyer |
| Charles B. Henderson | Dem. | NV | Feb-34 | Jul-47 | Former U.S. senator and lawyer |
| Hubert T. Stephens | Dem. | MS | Mar-35 | Feb-36 | Former U.S. senator and lawyer |
| Charles T. Fisher (Jr.) | Rep. | MI | Mar-35 | Dec-36 | Detroit banker |
| Emil Schram | Dem. | IN | Jun-36 | Jul-41 | Indiana farmer and irrigator |
| Howard J. Klossner | Rep. | MN | Apr-37 | Jul-45 | Minnesota banker |
| Sam H. Husbands | Dem. | SC | Aug-39 | Jan-46 | South Carolina banker |

| | Panel B – Determinants of RFC Board composition | | | |
|---|---|---|---|---|
| | All Boards | Excluding First | Political background | Business background |
| | (1) | (2) | (3) | (4) |
| Bank Capital in Default | 0.078 | -0.037 | -0.069 | 0.032 |
| | (0.530) | (0.690) | (0.410) | (0.326) |
| Banks per capita | 2.577 | 1.515 | -0.659 | 2.174* |
| | (0.577) | (0.495) | (0.606) | (0.096) |
| Log(Building permits per capita) | -0.019 | -0.028 | -0.003 | -0.025 |
| | (0.682) | (0.500) | (0.885) | (0.475) |
| Log(State Population) | 0.716 | 0.879 | 0.901* | -0.021 |
| | (0.418) | (0.141) | (0.061) | (0.928) |
| Log(Size per Railroad) | 0.773 | -0.452 | -0.465 | 0.013 |
| | (0.428) | (0.310) | (0.231) | (0.943) |
| Log(Number of Railroads) | 0.108 | -0.014 | -0.024* | 0.011 |
| | (0.185) | (0.536) | (0.090) | (0.507) |
| Railroad Bailout Weight | 0.102 | 0.142 | 0,985 | 0.047 |
| | (0.263) | (0.122) | (0.240) | (0.207) |
| Number of RFC directors | -0.159** | -0.008 | -0.113* | 0.105 |
| | (0.013) | (0.947) | (0.053) | (0.147) |
| $R^2$ | 0.179 | 0.058 | 0.084 | 0.094 |
| Observations | 344 | 344 | 344 | 344 |
| Year FE | Yes | Yes | Yes | Yes |
| State FE | Yes | Yes | Yes | Yes |

**Table 2: Summary statistics**

The sample consists of 1,928 annual observations for 183 railroads from 1927 to 1939. A bailout is defined as any loan from the Reconstruction Finance Corporation (RFC) or Public Works Administration (PWA) from 1932 to 1939. *Connections* is the number of states the railroad operated in that were homes to RFC directors in that year. *Leverage* is the ratio of long-term debt to total assets. *Bonds Due / T.A.* is the value of all bonds due between 1930 and 1934 to total assets in 1929. *Passenger / Total Revenue* equals passenger revenue divided by total revenue. *Volatility* is the standard deviation of the previous 12 month's earnings (if earnings was missing, the 12-month standard deviation of stock returns). *Average Wage* is the total expense on employees divided by the number of employees. We report tests of differences in means (t-test) and medians (Wilcoxon) between the groups. *, **, and *** denote significance at the 10%, 5% and 1% levels, respectively.

| | Panel A: Full sample | | | | | |
|---|---|---|---|---|---|---|
| | Bailed-out Railroads (N = 644 Railroad-Years) | | Non-Bailed-Out Railroads (N = 1,944 Railroad-Years) | | Difference | |
| | | | | | Mean | Median |
| | Mean | Median | Mean | Median | T-test | Wilcoxon |
| Log (Age, years) | 3.483 | 3.638 | 3.526 | 3.610 | -0.043 | 0.028 |
| Cash / T.A. | 0.015 | 0.013 | 0.023 | 0.014 | -0.008*** | -0.000 |
| Connections | 0.798 | 0.000 | 0.436 | 0.000 | -0.362*** | 0.000 |
| Log (Employment) | 8.980 | 9.121 | 7.223 | 7.102 | 1.757*** | 2.019*** |
| Leverage | 0.425 | 0.415 | 0.407 | 0.386 | 0.017 | 0.025*** |
| Loan size / T.A. | 0.006 | 0.000 | 0.000 | 0.000 | 0.006*** | 0.000 |
| Net income / T.A. | 0.009 | 0.009 | 0.013 | 0.000 | -0.004* | 0.008 |
| Passenger / Total Revenue | 0.116 | 0.096 | 0.099 | 0.055 | 0.018*** | 0.041*** |
| Log (Total Assets) | 19.089 | 19.176 | 17.193 | 16.996 | 1.897*** | 2.180*** |
| Volatility | 0.123 | 0.084 | 0.092 | 0.000 | 0.036*** | 0.084*** |
| Log (Average Wage) | 7.399 | 7.426 | 7.358 | 7.394 | 0.042*** | 0.032*** |

| | Panel B: Full sample | | | | | |
|---|---|---|---|---|---|---|
| | Multiple Bailouts (N = 532 Railroad-Years) | | One Bailout (N = 112 Railroad-Years) | | Difference | |
| | | | | | Mean | Median |
| | Mean | Median | Mean | Median | T-test | Wilcoxon |
| Log (Age, years) | 3.602 | 3.738 | 2.960 | 2.862 | 0.641*** | 0.876*** |
| Cash / T.A. | 0.014 | 0.012 | 0.019 | 0.016 | -0.006*** | 0.004*** |
| Connections | 0.883 | 0.000 | 0.393 | 0.000 | 0.491*** | 0.000 |
| Log (Employment) | 9.218 | 9.369 | 7.867 | 7.468 | 1.351*** | 1.901*** |
| Leverage | 0.419 | 0.419 | 0.451 | 0.407 | -0.032** | 0.012 |
| Loan size / T.A. | 0.011 | 0.000 | 0.003 | 0.000 | 0.008*** | 0.000 |
| Net income / T.A. | 0.009 | 0.001 | 0.006 | 0.002 | 0.003 | 0.001* |
| Passenger / Total Revenue | 0.125 | 0.101 | 0.076 | 0.064 | 0.049*** | 0.037*** |
| Log (Total Assets) | 19.325 | 19.302 | 17.991 | 17.537 | 1.334*** | 1.765*** |
| Volatility | 0.124 | 0.082 | 0.118 | 0.086 | 0.006 | 0.004 |
| Log (Average Wage) | 7.419 | 7.442 | 7.307 | 7.337 | 0.113*** | 0.005*** |

## Table 3: Determinants of bailout

We regress bailouts on railroad characteristics in a two-step regression. In step 1, Columns 1 and 2 present a regression where the dependent variable, *Application*, equals one if the railroad applied for at least one loan in that year, and zero otherwise. In step 2, we exclude all railroads that never applied for a bailout in Columns 3 to 6. Columns 3 and 4 present a regression where the dependent variable, *Approval*, equals one if the railroad applied for at least one loan in that year, and zero otherwise. Columns 5 and 6 present a regression where the dependent variable, *Approval Size*, is the railroad's approved bailout size divided by total assets. *Approval (in last 3 Years)* is a dummy variable that equals one if the railroad had an RFC or PWA loan approved in the last three years, and zero otherwise. *Cum. Loan Size / T.A.* equals the cumulative bailout loan amount a railroad has received since 1932 divided by its total assets. Other variables are defined in Table 2. All regressions include year- and region-fixed effects. *, **, and *** denote significance at the 10%, 5% and 1% levels, respectively.

|  | First step | | Second step | | | |
|---|---|---|---|---|---|---|
|  | *Application* | | *Approval* | | *Approval Size* | |
|  | (1) | (2) | (3) | (4) | (5) | (6) |
| Connections | 0.033*** | 0.029** | 0.028 | 0.038 | 0.000 | 0.000 |
|  | (0.005) | (0.041) | (0.195) | (0.162) | (0.846) | (0.832) |
| Log (Total Assets) | 0.044 | 0.060 | 0.202** | 0.242** | 0.004 | 0.003 |
|  | (0.268) | (0.113) | (0.048) | (0.037) | (0.503) | (0.720) |
| Net income / T.A. | -0.165 | -0.148 | -0.916 | -0.797 | -0.031 | -0.031 |
|  | (0.185) | (0.272) | (0.347) | (0.434) | (0.681) | (0.628) |
| Leverage | 0.018 | 0.049 | 0.283 | 0.697** | 0.019 | 0.064** |
|  | (0.694) | (0.360) | (0.337) | (0.038) | (0.233) | (0.025) |
| Cash / T. A. | 0.129 | 0.313 | 0.054 | 0.423 | 0.014 | -0.008 |
|  | (0.497) | (0.167) | (0.937) | (0.467) | (0.794) | (0.892) |
| Log (Age, years) | 0.019 | 0.015 | -0.086 | -0.116 | -0.002 | -0.005 |
|  | (0.702) | (0.778) | (0.351) | (0.206) | (0.691) | (0.232) |
| Volatility | -0.011 | -0.017 | 0.058 | 0.049 | 0.006 | 0.007 |
|  | (0.754) | (0.621) | (0.374) | (0.463) | (0.166) | (0.128) |
| Log (Employment) | -0.006 | -0.009 | -0.029 | -0.037 | -0.002 | -0.000 |
|  | (0.802) | (0.728) | (0.590) | (0.548) | (0.559) | (0.978) |
| Passenger / Total Revenue | 0.024 | 0.011 | 0.550** | 0.506* | -0.007 | 0.002 |
|  | (0.746) | (0.902) | (0.031) | (0.084) | (0.660) | (0.891) |
| Bonds Due$_{1930-1934}$ / T.A. | 0.174 | 0.172 | 1.079** | 0.853* | 0.024 | 0.018 |
|  | (0.258) | (0.310) | (0.011) | (0.071) | (0.365) | (0.534) |
| Approval (in Last 3 Years) |  |  | 0.061 | 0.047 | -0.005 | -0.006* |
|  |  |  | (0.307) | (0.421) | (0.253) | (0.091) |
| Cum. Loan size / T.A. |  |  | -0.935 | -1.483** | -0.121*** | -0.144*** |
|  |  |  | (0.146) | (0.018) | (0.000) | (0.001) |
| Adjusted $R^2$ | 0.283 | 0.232 | 0.218 | 0.125 | 0.075 | 0.032 |
| Freight Composition | Yes | Yes | Yes | Yes | Yes | Yes |
| Year FE | Yes | No | Yes | No | Yes | No |
| Firm FE | Yes | Yes | Yes | Yes | Yes | Yes |
| Rating FE | Yes | Yes | Yes | Yes | Yes | Yes |
| Region x Year FE | No | Yes | No | Yes | No | Yes |
| Observations | 1,694 | 1,694 | 619 | 619 | 615 | 615 |

## Table 4: Announcement effects

We calculate the abnormal returns (AR) and cumulative abnormal returns (CAR) of a security from four days before to four days after the announcement of an application, approval, or refusal. We measure AR as the security's returns less the Moody's bond index / CRSP market index on the same day. We average the AR (CAR) across securities. Standard errors are clustered by railroads. *p*-values appear in parentheses. Panel A presents the results for all bailouts. Panel B presents the results for the initial and subsequent bailouts. *, **, and *** denote significance at the 10%, 5% and 1% levels, respectively.

| | Panel A: All bailouts | | | | | |
|---|---|---|---|---|---|---|
| | *Applications* | | *Approvals* | | *Refusals* | |
| AR | Bonds | Equity | Bonds | Equity | Bonds | Equity |
| Day -4 | -0.001 | -0.007 | -0.003 | 0.005 | -0.015 | 0.048 |
| | (0.719) | (0.237) | (0.886) | (0.289) | (0.199) | (0.320) |
| Day -3 | 0.002 | 0.009 | 0.003 | -0.003 | -0.008 | 0.002 |
| | (0.515) | (0.181) | (0.326) | (0.548) | (0.692) | (0.916) |
| Day -2 | -0.001 | 0.002 | 0.001 | 0.005 | -0.003 | -0.032 |
| | (0.752) | (0.681) | (0.709) | (0.397) | (0.905) | (0.174) |
| Day -1 | 0.005 | -0.003 | -0.003 | 0.006 | 0.011 | -0.079* |
| | (0.201) | (0.969) | (0.222) | (0.419) | (0.648) | (0.092) |
| Day 0 | 0.009* | 0.004 | 0.003 | -0.025 | -0.019 | 0.044 |
| | (0.068) | (0.550) | (0.278) | (0.516) | (0.340) | (0.443) |
| Day +1 | 0.003 | 0.006 | 0.007*** | -0.008 | -0.051* | -0.061* |
| | (0.239) | (0.302) | (0.003) | (0.161) | (0.085) | (0.100) |
| Day +2 | 0.003 | -0.001 | 0.003 | 0.002 | -0.029 | 0.027 |
| | (0.242) | (0.922) | (0.326) | (0.698) | (0.434) | (0.121) |
| Day +3 | 0.003 | 0.003 | 0.006* | -0.009 | 0.021 | -0.029 |
| | (0.260) | (0.646) | (0.100) | (0.175) | (0.229) | (0.337) |
| Day +4 | -0.000 | -0.005 | -0.000 | 0.007 | 0.028 | 0.005 |
| | (0.993) | (0.263) | (0.986) | (0.215) | (0.216) | (0.749) |
| CAR | 0.022*** | 0.011 | 0.018** | -0.017 | -0.066 | -0.075 |
| (t-4, t+4) | (0.009) | (0.280) | (0.028) | (0.622) | (0.329) | (0.247) |
| Observations | 196 | 135 | 240 | 191 | 9 | 9 |

|  | Panel B: First vs. subsequent bailouts | | | | | | | |
|---|---|---|---|---|---|---|---|---|
|  | First bailout | | | | Subsequent bailouts | | | |
|  | *Applications* | | *Approval* | | *Applications* | | *Approval* | |
| AR | Bonds | Equity | Bonds | Equity | Bonds | Equity | Bonds | Equity |
| Day -4 | -0.006 | 0.004 | -0.007 | 0.001 | 0.001 | -0.010 | 0.001 | 0.006 |
|  | (0.134) | (0.731) | (0.203) | (0.942) | (0.660) | (0.159) | (0.532) | (0.248) |
| Day -3 | 0.001 | 0.029** | -0.001 | 0.013 | 0.002 | 0.004 | 0.003 | -0.007 |
|  | (0.769) | (0.250) | (0.834) | (0.537) | (0.563) | (0.631) | (0.276) | (0.221) |
| Day -2 | 0.001 | -0.033*** | 0.008 | 0.003 | -0.002 | 0.013* | -0.000 | 0.006 |
|  | (0.924) | (0.001) | (0.533) | (0.862) | (0.503) | (0.069) | (0.887) | (0.394) |
| Day -1 | 0.011 | 0.004 | -0.002 | -0.012 | 0.003 | -0.001 | -0.003 | 0.009 |
|  | (0.402) | (0.785) | (0.649) | (0.575) | (0.234) | (0.860) | (0.257) | (0.248) |
| Day 0 | 0.021* | 0.001 | 0.012** | -0.229 | 0.004* | 0.005 | 0.000 | 0.015* |
|  | (0.084) | (0.931) | (0.038) | (0.317) | (0.086) | (0.505) | (0.926) | (0.089) |
| Day +1 | 0.004 | -0.007 | 0.011* | 0.001 | 0.003 | 0.011 | 0.006** | -0.010* |
|  | (0.521) | (0.591) | (0.060) | (0.974) | (0.318) | (0.125) | (0.019) | (0.084) |
| Day +2 | 0.004* | 0.029*** | 0.003 | 0.011 | 0.002 | -0.009 | 0.003 | 0.003 |
|  | (0.096) | (0.008) | (0.638) | (0.461) | (0.449) | (0.211) | (0.388) | (0.995) |
| Day +3 | 0.004 | -0.009 | 0.029* | -0.037* | 0.002 | 0.007 | -0.000 | -0.003 |
|  | (0.304) | (0.559) | (0.092) | (0.093) | (0.454) | (0.307) | (0.993) | (0.618) |
| Day +4 | 0.002 | -0.001 | 0.007* | 0.012 | -0.001 | -0.006 | -0.002 | 0.006 |
|  | (0.656) | (0.861) | (0.065) | (0.472) | (0.779) | (0.302) | (0.512) | (0.309) |
| CAR | 0.041** | 0.017 | 0.059** | -0.239 | 0.014* | 0.013 | 0.009 | 0.023* |
| (t-4, t+4) | (0.050) | (0.563) | (0.031) | (0.302) | (0.071) | (0.362) | (0.288) | (0.081) |
| Observations | 56 | 32 | 47 | 32 | 141 | 103 | 193 | 159 |

## Table 5: Determinants of announcement CARs

We regress a railroad's cumulative abnormal bond or equity return from four days before to four days after an application or approval of an RFC or PWA loan. Variables are as defined in Table 2. We add region, year and rating fixed effects and cluster standard errors at the railroad level. *, **, and *** denote significance at the 10%, 5% and 1% levels, respectively.

|  | Application | | Approval | |
|---|---|---|---|---|
| Security | Bond | Equity | Bond | Equity |
|  | (1) | (2) | (3) | (4) |
| Connections | -0.001 | 0.015 | -0.008 | 0.062 |
|  | (0.976) | (0.679) | (0.470) | (0.113) |
| Log (Total Assets) | -0.045 | 5.110*** | -0.092 | 2.621 |
|  | (0.986) | (0.000) | (0.396) | (0.356) |
| Net income / T.A. | 1.019 | 1.524 | 1.005 | 7.484 |
|  | (0.588) | (0.536) | (0.163) | (0.227) |
| Leverage | -0.663* | -2.639*** | -0.149 | -6.578 |
|  | (0.098) | (0.000) | (0.536) | (0.160) |
| Cash / T.A. | -13.559* | 8.353 | 9.211*** | 21.801* |
|  | (0.076) | (0.188) | (0.004) | (0.081) |
| Log (Age, years) | -0.037 | 0.006 | -0.006* | -0.124* |
|  | (0.153) | (0.806) | (0.062) | (0.100) |
| Volatility | 0.186 | 0.079 | -0.002 | -0.027 |
|  | (0.190) | (0.308) | (0.787) | (0.756) |
| Log (Employment) | -0.094 | -2.748*** | 0.142 | 0.879 |
|  | (0.918) | (0.000) | (0.508) | (0.655) |
| Passenger / Total Revenue | 0.264*** | 0.527*** | -0.123* | -3.133* |
|  | (0.001) | (0.008) | (0.072) | (0.091) |
| Loan size / T.A. |  |  | 0.122 | -3.704 |
|  |  |  | (0.660) | (0.167) |
| Decision time |  |  | -0.000 | -0.000 |
|  |  |  | (0.799) | (0.996) |
| Loan size difference |  |  | -0.190 | -0.316 |
|  |  |  | (0.220) | (0.909) |
| Firm FE | Yes | Yes | Yes | Yes |
| Rating FE | Yes | Yes | Yes | Yes |
| Year FE | Yes | Yes | Yes | Yes |
| Freight composition | Yes | Yes | Yes | Yes |
| Observations | 195 | 141 | 231 | 180 |
| $R^2$ | 0.072 | 0.103 | 0.053 | 0.010 |

## Table 6: Determinants of bond defaults

We regress bond defaults on lagged railroad characteristics. *Default* equals one if the railroad failed to meet a coupon or principal repayment, or in any way changed the terms of the issue in the current year. We drop all observations of the respective railroads the year after *Default* equals one. *Approval* equals one if the railroad obtained an RFC or PWA loan in the previous year. In columns 1 and 4, we run a probit regression model. In column 2 and 5, we present our first-stage regression for the instrumental-variable (IV) approach. We regress an indicator variable equal to one in the year the railroad received an *Approval*, and zero otherwise, on railroad controls. In columns 3 and 6, we present the second-stage instrumental variable (IV) regression. *p*-values, in parentheses, are adjusted for heteroskedasticity and clustered at the railroad-level. We include region, year and (in columns 4-6) bond rating fixed effects. For a railroad with multiple outstanding bonds, we use the rating of the bond closest to maturity. *, **, and *** denote significance at the 10%, 5% and 1% levels, respectively.

| | Bond default (this year) | | | Bond default (this year or next two years) | | |
|---|---|---|---|---|---|---|
| | | First Stage | Second Stage | | First Stage | Second Stage |
| | (1) | (2) | (3) | (4) | (5) | (6) |
| Approval | 0.441 | | -0.834 | 0.756*** | | 1.269 |
| | (0.330) | | (0.796) | (0.006) | | (0.578) |
| Log (Total Assets) | 0.262 | 0.016 | 0.272 | 0.348* | 0.017 | 0.339 |
| | (0.295) | (0.180) | (0.262) | (0.085) | (0.140) | (0.140) |
| Net income / T.A. | -9.260*** | -0.213 | -9.233*** | -6.049* | -0.193 | -5.912 |
| | (0.009) | (0.130) | (0.006) | (0.100) | (0.104) | (0.132) |
| Leverage | 0.444 | 0.032 | 0.473 | 0.597 | 0.035 | 0.578 |
| | (0.289) | (0.296) | (0.242) | (0.186) | (0.198) | (0.182) |
| Cash / T.A. | -80.589*** | 0.283** | -76.855*** | -38.876** | 0.294*** | -38.911** |
| | (0.000) | (0.038) | (0.007) | (0.011) | (0.008) | (0.011) |
| Log (Age, years) | -0.531*** | -0.022 | -0.539*** | -0.449*** | -0.004 | -0.445*** |
| | (0.004) | (0.179) | (0.005) | (0.005) | (0.711) | (0.006) |
| Volatility | 0.321 | -0.048** | 0.265 | 0.270 | -0.0144 | 0.278 |
| | (0.254) | (0.046) | (0.443) | (0.160) | (0.682) | (0.161) |
| Log (Employment) | -0.317 | 0.016 | -0.275 | -0.375* | 0.009 | -0.382** |
| | (0.183) | (0.132) | (0.272) | (0.059) | (0.321) | (0.043) |
| Bonds Due$_{1930-1934}$ / T.A. | 0.820 | 0.068 | 0.946 | 1.269 | 0.002 | 1.256 |
| | (0.572) | (0.616) | (0.517) | (0.346) | (0.977) | (0.344) |
| Passenger / Total Revenue | -1.865 | -0.059 | -0.275 | -0.730 | -0.102* | -0.667 |
| | (0.461) | (0.271) | (0.272) | (0.654) | (0.063) | (0.678) |
| Cum. Loan Size / T.A. | -1.037 | 3.710*** | 3.961 | -0.036 | 3.871*** | -2.149 |
| | (0.686) | (0.000) | (0.761) | (0.991) | (0.000) | (0.819) |
| Connections | | 0.042*** | | | 0.046*** | |
| | | (0.002) | | | (0.001) | |
| Region FE | Yes | Yes | Yes | Yes | Yes | Yes |
| Year FE | Yes | Yes | Yes | Yes | Yes | Yes |
| Rating FE | Yes | Yes | Yes | Yes | Yes | Yes |
| Freight Composition | Yes | Yes | Yes | Yes | Yes | Yes |
| Log Likelihood | -67.736 | -52.346 | -52.346 | -189.601 | -88.920 | -88.920 |
| Observations | 847 | 847 | 847 | 1,217 | 1,217 | 1,217 |
| F-Statistic | n.a. | 36.370 | n.a. | n.a. | 36.370 | n.a |

## Table 7: Rating changes

We regress bond rating changes, for the nearest-to-maturity bond, on lagged railroad and bond characteristics. Columns 1 and 4 contain probit regressions. The dependent variable equals one if there was a Moody's rating downgrade in the current year, and zero otherwise (Column 1) or if there was more than one rating downgrade from the current year to year t+2 (Column 4). We drop all observations of the respective railroads the year after *Default* equals one. Variables are as defined in Tables 2 and 6. We include *Time to maturity*, the log of the number of years to maturity for the respective bond, and the *Nominal outstanding amount of the bond to total long-term debt*. In columns 2 and 5, we report first-stage regressions. We regress an indicator variable equal to one the year the railroad received an *Approval*, and zero otherwise, on railroad and bond controls. We present the second-stage instrumental variable (IV) probit regressions for one downgrade (column 3) and multiple downgrades (column 6). *p*-values, in parentheses, are adjusted for heteroskedasticity and clustered at the railroad-level. We include region, firm, and rating fixed effects. *, **, and *** denote significance at the 10%, 5% and 1% levels, respectively.

|  | Single Rating Downgrade (this year) | | | Multiple Rating Downgrades (this year or next two years) | | |
|---|---|---|---|---|---|---|
|  | Probit | First Stage | Second Stage | Probit | First Stage | Second Stage |
|  | (1) | (2) | (3) | (4) | (5) | (6) |
| Approval | -0.351** |  | -0.702* | -1.173*** |  | -3.203*** |
|  | (0.021) |  | (0.094) | (0.000) |  | (0.000) |
| Connections |  | 0.082*** |  |  | 0.085*** |  |
|  |  | (0.000) |  |  | (0.000) |  |
| **Railroad control variables** | | | | | | |
| Log (Total Assets) | -1.559 | 0.259 | -1.469 | -1.644 | 0.359** | 0.055 |
|  | (0.295) | (0.253) | (0.312) | (0.257) | (0.044) | (0.952) |
| Net income / T.A. | 11.981*** | -2.621*** | 10.485** | 2.894 | -0.241 | -0.509 |
|  | (0.003) | (0.002) | (0.025) | (0.520) | (0.104) | (0.603) |
| Leverage | -9.321*** | 1.133*** | -8.828*** | -7.112*** | 0.907*** | -1.969 |
|  | (0.000) | (0.004) | (0.000) | (0.002) | (0.007) | (0.296) |
| Cash / T.A. | -38.059*** | 5.041*** | -35.721*** | -2.132 | 1.611* | 2.691 |
|  | (0.000) | (0.003) | (0.000) | (0.834) | (0.064) | (0.671) |
| Log (Age, years) | -0.006 | -0.411** | -0.089 | -7.289*** | -0.462*** | -5.317*** |
|  | (0.993) | (0.012) | (0.900) | (0.000) | (0.002) | (0.000) |
| Volatility | 0.159 | 0.108*** | 0.194 | -1.026 | 0.080*** | -0.333 |
|  | (0.241) | (0.000) | (0.180) | (0.117) | (0.000) | (0.357) |
| Log (Employment) | 3.739*** | -0.499*** | 3.495*** | 3.657*** | -0.407*** | 0.901* |
|  | (0.000) | (0.000) | (0.000) | (0.000) | (0.000) | (0.061) |
| Passenger / Total Revenue | 20.885*** | -1.357* | 19.672*** | 22.443*** | -0.553 | 9.146** |
|  | (0.000) | (0.093) | (0.000) | (0.001) | (0.437) | (0.024) |
| Cum. Loan Size / T.A. | 4.707 | 3.563*** | 6.147 | 0.950 | 3.493*** | 11.351*** |
|  | (0.187) | (0.000) | (0.154) | (0.732) | (0.000) | (0.000) |
| **Bond control variables** | | | | | | |
| Time to Maturity | -0.081* | 0.013 | -0.077* | -0.057 | 0.0145* | -0.002 |
|  | (0.090) | (0.113) | (0.098) | (0.362) | (0.056) | (0.952) |
| Nominal Outstanding / L.T.D. | -6.975 | 0.325 | -6.774 | -4.616 | -0.058 | -3.310 |
|  | (0.419) | (0.539) | (0.421) | (0.224) | (0.709) | (0.179) |
| Region FE | Yes | Yes | Yes | Yes | Yes | Yes |
| Firm FE | Yes | Yes | Yes | Yes | Yes | Yes |
| Rating FE | Yes | Yes | Yes | Yes | Yes | Yes |
| Freight Composition | Yes | Yes | Yes | Yes | Yes | Yes |
| Log likelihood | -657.589 | -1,063.879 | -1,063.879 | -524,494 | -899.982 | -899.982 |
| Observations | 1,419 | 1,419 | 1,419 | 1,645 | 1,645 | 1,645 |
| F-Statistic | n.a. | 100.885 | n.a. | n.a. | 100.885 | n.a. |
| Specification | Probit | IV-Probit | IV-Probit | Probit | IV-Probit | IV-Probit |

**Table 8: Difference-in-difference: Leverage, employment, wages, and profitability**

We present difference-in-difference regressions for leverage, log(employment), log(average wage), and profitability, with variation in treatment timing and multiple periods following Callaway and Sant'Anna (2021). In Panel A, *ATT* is defined as the average treatment effect for the treated population in the years following the treatment. Panel B presents the results of an event study analysis. Panel C presents results if we assume that bailed-out railroads (counterfactually) received an RFC or PWA loan in 1929, and the post-bailout-period was 1929-32. *p*-values in parentheses use doubly robust standard errors, following Sant'Anna and Zhao (2020). Chi-squared is the chi-squared pretrend test. All regressions use year and region-fixed effects. *, **, and *** denote significance at the 10%, 5% and 1% levels, respectively.

|  | Panel A: Average treatment | | | |
|---|---|---|---|---|
|  | *Leverage* | *Employment* | *Average Wage* | *Profitability* |
|  | (1) | (2) | (3) | (4) |
| ATT | -0.026* | 0.047 | 0.045* | 0.001 |
|  | (0.060) | (0.137) | (0.100) | (0.845) |
| Chi-squared | 19.971 | 7.162 | 9.362 | 3.910 |
|  | (0.523) | (0.613) | (0.326) | (0.750) |
| Obs. | 1,838 | 1,838 | 1,838 | 1,838 |
|  | Panel B: Event study | | | |
| Year - 4 | 0.007 | 0.054 | 0.060 | -0.003 |
|  | (0.318) | (0.334) | (0.771) | (0.500) |
| Year - 3 | -0.003 | -0.009 | -0.185 | -0.002 |
|  | (0.625) | (0.561) | (0.395) | (0.566) |
| Year -2 | 0.011* | -0.013 | 0.123 | -0.003 |
|  | (0.090) | (0.338) | (0.325) | (0.436) |
| Year -1 | -0.011 | -0.013 | -0.022 | -0.004 |
|  | (0.250) | (0.231) | (0.276) | (0.206) |
| Year 0 | 0.004 | 0.016 | 0.036* | -0.011 |
|  | (0.489) | (0.217) | (0.085) | (0.294) |
| Year +1 | -0.010 | 0.045** | 0.044** | 0.006 |
|  | (0.298) | (0.040) | (0.048) | (0.192) |
| Year +2 | -0.015 | 0.069** | 0.049* | -0.000 |
|  | (0.294) | (0.039) | (0.061) | (0.984) |
| Year +3 | -0.038** | 0.055 | 0.051* | 0.002 |
|  | (0.042) | (0.144) | (0.058) | (0.766) |
| Year +4 | -0.023 | 0.055 | 0.043 | 0.006 |
|  | (0.234) | (0.151) | (0.136) | (0.403) |
| Obs. | 1,838 | 1,838 | 1,838 | 1,838 |
|  | Panel C: Placebo test | | | |
| ATT | 0.006 | 0.027 | -0.001 | -0.008 |
|  | (0.671) | (0.625) | (0.827) | (0.194) |
| Obs. | 913 | 913 | 913 | 913 |

**Table 9: Instrumental variable regressions: Leverage, employment, wages, and profitability**

In the first stage, we regress *Approval* on *Connections* and lagged railroad characteristics. In the second stage, we regress contemporaneous railroad leverage, log(employment), log(average wage), and profitability on the fitted level of lagged *Approval* and lagged characteristics. Variables are as defined in Table 2 and 6. *p*-values are adjusted for heteroskedasticity and clustered at the firm-level, in parentheses. *, **, and *** denote significance at the 10%, 5% and 1% levels, respectively.

|  | First-Stage | Second-Stage | | | |
|---|---|---|---|---|---|
|  |  | *Leverage* | *Employment* | *Average Wage* | *Profitability* |
|  | (1) | (2) | (3) | (4) | (5) |
| Approval |  | -0.102* | -0.081 | 0.154* | 0.015 |
|  |  | (0.058) | (0.628) | (0.073) | (0.291) |
| Log (Total Assets) | 0.074* | 0.032 | 0.027 | 0.019 | 0.006 |
|  | (0.067) | (0.126) | (0.360) | (0.340) | (0.271) |
| Net income / T.A. | 0.153 | -0.021 | -0.099 | 0.076 | 0.435*** |
|  | (0.243) | (0.738) | (0.661) | (0.540) | (0.000) |
| Leverage | 0.043 | 0.591*** | 0.079 | 0.067 | 0.019* |
|  | (0.399) | (0.000) | (0.357) | (0.296) | (0.051) |
| Cash / T. A. | -0.059 | 0.041 | 0.095 | -0.030 | 0.049 |
|  | (0.830) | (0.625) | (0.699) | (0.882) | (0.341) |
| Log (Age, years) | 0.036 | -0.009 | 0.108*** | -0.014 | 0.001 |
|  | (0.593) | (0.426) | (0.007) | (0.519) | (0.804) |
| Volatility | -0.015 | -0.012* | -0.002 | 0.011 | -0.005** |
|  | (0.609) | (0.078) | (0.905) | (0.243) | (0.034) |
| Log (Employment) | -0.028 | -0.014 | 0.414*** | -0.049** | 0.001 |
|  | (0.307) | (0.448) | (0.000) | (0.040) | (0.738) |
| Passenger / Total Revenue | -0.013 | -0.038 | -0.023 | 0.035 | -0.013** |
|  | (0.870) | (0.370) | (0.677) | (0.432) | (0.017) |
| Bonds Due$_{1930-1934}$ / T.A. | -0.050 | -0.033 | -0.108 | 0.032 | -0.022 |
|  | (0.669) | (0.676) | (0.512) | (0.506) | (0.113) |
| Connections | 0.059*** |  |  |  |  |
|  | (0.000) |  |  |  |  |
| Firm FE | Yes | Yes | Yes | Yes | Yes |
| Year FE | Yes | Yes | Yes | Yes | Yes |
| Region FE | Yes | Yes | Yes | Yes | Yes |
| Freight Composition | Yes | Yes | Yes | Yes | Yes |
| $R^2$ | 0.179 | 0.803 | 0.955 | 0.002 | 0.295 |
| F statistic | 76.140 | n.a. | n.a. | n.a. | n.a. |
| Observations | 1,645 | 1,771 | 1,771 | 1,771 | 1,816 |

**Table 10: Railroad bailouts and building approvals**

We regress the logarithm of building permits per city on *City RFC Approvals* (the fraction of all railroads that pass through the city that received an RFC/PWA railroad loan approval the previous year). We condition on state-level bank characteristics: the logarithm of bank loans per capita; the logarithm of bank deposits per capita; the logarithm of the number of all banks; and the capital of nationally chartered banks that operated in the state that were liquidated in year *t* divided by the capital of all nationally-chartered banks in that state in year *t*. We instrument *City RFC Approvals* with *Connections*. *p*-values are adjusted for heteroskedasticity and clustered at the city-level, in parentheses. *, **, and *** denote significance at the 10%, 5% and 1% levels, respectively.

|  | First Stage | Second Stage | First Stage | Second Stage | First Stage | Second Stage |
|---|---|---|---|---|---|---|
|  | (1) | (2) | (3) | (4) | (5) | (6) |
| City RFC Approvals |  | -1.549*** |  | -1.413*** |  | -0.293 |
|  |  | (0.000) |  | (0.000) |  | (0.239) |
| Log (Loans per Capita) | -0.126*** | 0.221*** | -0.058 | 0.142 | 0.323*** | -0.151 |
|  | (0.000) | (0.004) | (0.183) | (0.149) | (0.001) | (0.319) |
| Log (Deposits per Capita) | -0.266*** | 1.686*** | -0.291*** | 2.725*** | 0.176 | 0.339 |
|  | (0.000) | (0.000) | (0.008) | (0.000) | (0.115) | (0.161) |
| Log (Number of Banks) | 0.152*** | -0.815*** | 0.075 | -1.516*** | -0.489*** | 0.069 |
|  | (0.000) | (0.000) | (0.451) | (0.000) | (0.000) | (0.765) |
| Capital of Suspended Banks | 0.747*** | 1.569*** | 0.709*** | 0.969** | 0.420* | 0.746* |
|  | (0.002) | (0.005) | (0.005) | (0.043) | (0.073) | (0.048) |
| Railroad Size | -0.002 | -0.016 | 0.414 | -2.461*** | -0.209 | -0.892* |
|  | (0.973) | (0.882) | (0.127) | (0.003) | (0.530) | (0.070) |
| Log (Number of Railroads) | -0.156*** | 0.183 |  |  |  |  |
|  | (0.000) | (0.302) |  |  |  |  |
| Connections | 0.258*** |  | 0.253*** |  | 0.118*** |  |
|  | (0.000) |  | (0.000) |  | (0.000) |  |
| Year FE | No | No | No | No | Yes | Yes |
| City FE | No | No | Yes | Yes | Yes | Yes |
| R² | 0.016 | 0.118 | 0.019 | 0.042 | 0.072 | 0.019 |
| Observations | 2,026 | 2,231 | 2,026 | 2,231 | 2,026 | 2,321 |
| F-statistic | 122.543 | n.a. | 115.479 | n.a. | 31.4 | n.a. |

**Table 11: Announcement effects for related firms**

We calculate the abnormal return (AR) and cumulative abnormal return (CAR) for related firms' equity after the announcement of a railroad's bailout application (Panel A) and approval (Panel B). We measure the AR as a firm's equity return less the CRSP market index. An overlap of *Yes* indicates the railroad/manufacturing firm operates in at least one city with the bailed-out railroad. An overlap of *No* indicates the railroad/manufacturing firm does not operate in any cities in which the bailed-out railroad operates. *High* indicates that the percentage overlap is above the mean level across all firms. *Low* indicates that the percentage overlap is non-zero and below the mean overlap across all firms. For railroads, the percentage overlap is defined as the number of cities that both railroads serve divided by the total number of cities of the bailed-out railroad. For manufacturing firms, the percentage overlap is defined as the number of cities that the railroad and manufacturing firm both operate in divided by the total number of cities the manufacturer operates in. *p*-values appear in parentheses. We report the *p*-values of t-test differences between the groups (*Yes – No* or *High - Low*) in Diff. *, **, and *** denote significance at the 10%, 5% and 1% levels, respectively. The returns are winsorized at the 2.5% level.

|  | Railroads | | | | Manufacturing | | | |
|---|---|---|---|---|---|---|---|---|
| Overlap | Yes | No | High | Low | Yes | No | High | Low |
|  | (1) | (2) | (3) | (4) | (5) | (6) | (7) | (8) |
| | | | | Panel A: Application | | | | |
| Day -1 | 0.003** | 0.002* | 0.004** | 0.002 | 0.003*** | 0.002*** | 0.004*** | 0.002*** |
|  | (0.045) | (0.055) | (0.042) | (0.574) | (0.000) | (0.000) | (0.001) | (0.005) |
| Day 0 | 0.005*** | 0.003** | 0.006** | 0.003 | 0.000 | -0.000** | 0.000 | 0.000 |
|  | (0.004) | (0.032) | (0.016) | (0.120) | (0.647) | (0.024) | (0.781) | (0.971) |
| Day +1 | 0.006*** | 0.007*** | 0.006*** | 0.007*** | 0.003*** | 0.001*** | 0.005*** | 0.002* |
|  | (0.000) | (0.000) | (0.009) | (0.000) | (0.000) | (0.000) | (0.000) | (0.059) |
| CAR | 0.015*** | 0.012*** | 0.016*** | 0.014*** | 0.006*** | 0.005*** | 0.009*** | 0.002*** |
| (t-1, t+1) | (0.000) | (0.000) | (0.000) | (0.000) | (0.000) | (0.000) | (0.000) | (0.005) |
| Diff | 0.003 | | 0.003 | | 0.001* | | 0.006*** | |
| (t-1, t+1) | [0.895] | | [0.385] | | [1.872] | | [3.791] | |
| Obs. | 2,217 | 4,931 | 1,451 | 756 | 12,907 | 68,865 | 5,326 | 7,759 |
| | | | | Panel B: Approval | | | | |
| Day -1 | -0.001 | -0.001 | -0.000 | -0.001 | 0.003*** | 0.001*** | 0.003*** | 0.002*** |
|  | (0.355) | (0.351) | (0.814) | (0.223) | (0.000) | (0.000) | (0.000) | (0.000) |
| Day 0 | 0.002 | 0.004** | 0.002 | 0.002 | 0.001*** | 0.002*** | 0.002** | 0.008 |
|  | (0.262) | (0.027) | (0.401) | (0.482) | (0.007) | (0.000) | (0.030) | (0.143) |
| Day +1 | 0.003 | 0.003** | -0.002 | 0.005* | 0.001*** | -0.001*** | 0.003** | 0.008 |
|  | (0.842) | (0.012) | (0.136) | (0.082) | (0.005) | (0.008) | (0.010) | (0.111) |
| CAR | 0.001 | 0.005*** | -0.001 | 0.005*** | 0.005*** | 0.004*** | 0.007*** | 0.004*** |
| (t-1, t+1) | (0.653) | (0.000) | (0.631) | (0.025) | (0.000) | (0.000) | (0.000) | (0.000) |
| Diff | -0.005* | | -0.006** | | 0.002** | | 0.003** | |
| (t-1, t+1) | [-1.695] | | [-1.921] | | [2.340] | | [2.436] | |
| Obs. | 2,939 | 6,191 | 1,923 | 1,001 | 17,475 | 70,600 | 6,705 | 10,770 |